\documentclass[11pt]{article}
\usepackage{moriond,epsfig}

\def\NPB{{\em Nucl. Phys.} B}
\def\PLB{{\em Phys. Lett.}  B}

\def\PRD{{\em Phys. Rev.} D}
\def\PRC{{\em Phys. Rev.} C}

\def\be{\begin{equation}}
\def\ee{\end{equation}}
\def\bea{\begin{eqnarray}}
\def\eea{\end{eqnarray}}

\usepackage{graphicx} 

\newcommand{\trD}[1]{\mbox{\boldmath $#1$}}

\newcommand{\vers}[1]{\hat{\trD{#1}}}

\newcommand{\spub}[2]{\overline{u}_{#1}(#2)}
\newcommand{\spv}[2]{v_{#1}(#2)}

\begin{document}

\title{CHIRAL SYMMETRY BREAKING WITH SCALAR CONFINEMENT}

\author{ P. Bicudo and G. Marques }

\address{Dep. F\'{\i}sica and CFIF, Instituto Superior T\'ecnico, Av. 
Rovisco Pais, 1049-001 Lisboa, Portugal}

\maketitle
\abstracts{
Spontaneous chiral symmetry breaking is accepted
to occur in low energy hadronic physics, resulting in the
several successful theorems of PCAC. On the other hand
scalar confinement is suggested both by the spectroscopy
of hadrons and by the string picture of confinement.
However these two evidences are apparently conflicting,
because chiral symmetry breaking requires a chiral invariant
coupling to the quarks, say a vector coupling like in QCD.
Here we reformulate the coupling of the quarks to the string,
and we are able to comply with chiral symmetry breaking, using
scalar confinement. The results are quite encouraging. 
}

\section{Open Problem}

Recently Bjorken asked,
'' how are the many disparate methods of 
describing hadrons which are now in use related to each other 
and to the first principles of QCD?''.
Chiral symmetry breaking and scalar confinement 
are apparently conflicting,
because chiral symmetry breaking requires a chiral invariant
coupling to the quarks, say a vector coupling like in QCD.
Here we try to solve this old conflict of hadronic physics, 
which remained open for many years. 

The QCD Lagrangian is chiral invariant in the 
limit of vanishing quark masses. This is crucial because 
Spontaneous chiral symmetry breaking is accepted to occur in 
low energy hadronic physics, for the light flavors u, d and  s, where,
$mu , md << ms < \Lambda_{QCD} < M_N/3$.
The techniques of current algebra led to 
beautifully correct theorems, the  PCAC (Partially Conserved
Axial Current) theorems. 
The QM (Quark Models) are widely used as a simplification of QCD, 
convenient to study quark bound states and hadron scattering. 
Recently 
\cite{Bicudo} 
we have shown that these beautiful PCAC theorems, 
like the Weinberg theorem for $\pi-\pi$ scattering, are reproduced by quark 
models with spontaneous chiral symmetry breaking. 

On the other hand the confining potential for constituent quarks is probably scalar.
Scalar confinement is suggested both by the spectroscopy
of hadrons, by lattice simulations and by the string picture of confinement.
In a pertubative QCD scenario, the hadron spectroscopy would be qualitatively 
similar to electronic spectra of the lighter atoms. 
It is remarkable that the Spin-Orbit potential (also called fine interaction 
in atomic physics) turns out to be suppressed in hadronic spectra because it is 
smaller than the Spin-Spin potential (also called hyperfine interaction).
This constitutes an evidence of non-pertubative QCD. Another evidence of
non-pertubative QCD is present in the angular and radial excitations of hadrons,
which fit linear trajectories in Regge plots, and suggest a long range,
probably linear, confining potential for the quarks. 
This led Henriques, Kellet and Moorhouse, Isgur and Karl, and others  
\cite{Henriques,Isgur}
to develop a Quark Model 
where a short-range vector potential plus a long-range scalar potential 
partly cancel the Spin-Orbit interaction. The short range potential 
is inspired in the one gluon exchange, and the quark vertex is a Coulomb-like
potential, with a vector coupling $\bar \psi \gamma^\mu \psi$. The long range 
potential has  scalar coupling $\bar \psi \psi$, and is a linear potential.

\begin{figure}
\includegraphics{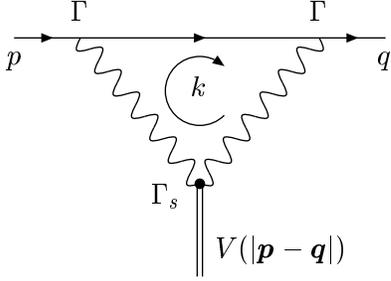}
\label{loopvertex} 
\caption{The coupling of a quark to a string with a double gluon vertex} 
\end{figure}

In this talk we define a quark model that matches the apparently conflicting vector 
coupling of QCD with a scalar confinement. We also solve the mass gap equation for the 
dynamical generation of the quark mass with the spontaneous breaking of chiral symmetry,
and we indeed generate both the constituent quark mass and the scalar confinement.

\section{The double vertex non-pertubative confining interaction}
\label{sec:potential}

We aim to couple a quark line in a Feynman diagram with a 
scalar string, using the vector gluon-quark coupling of QCD. We remark that the a simple 
vertex does not solve this problem, therefore we reformulate the standard 
coupling of the quark to the confining potential. The coupling needs at least a 
double vertex, similar to the vertices that couple a quark to a gluon ladder in 
models of the pomeron. Our double vertex is depicted in Fig. \ref{loopvertex}. 

To get the Dirac coupling of each gluon to a light quark,
we follow the coupling obtained in the heavy-light quark system, computed
in the local coordinate gauge
\cite{Nora}. This results in a Dirac coupling a pair of 
$\gamma^0$ matrices, which is also compatible with the Coulomb gauge.

In the color sector, each sub-vertex couples with the Gell-Mann 
matrix $\lambda^a /2$. Moreover the string is also a colored object because
it contains the flux of color electric field. In quark models the string usually
couples with a $\lambda^a /2$ to the quark line, here it couples with to the
two $\lambda^a /2$ of the sub-vertices. For a scalar coupling, which is
symmetric, we use the symmetric structure function $d^{abc}$ defined with,
\begin{equation}
\{ \lambda^a ,\lambda^b \}= d^{abc}\lambda_c \ . 
\end{equation}

\begin{figure} 
\includegraphics{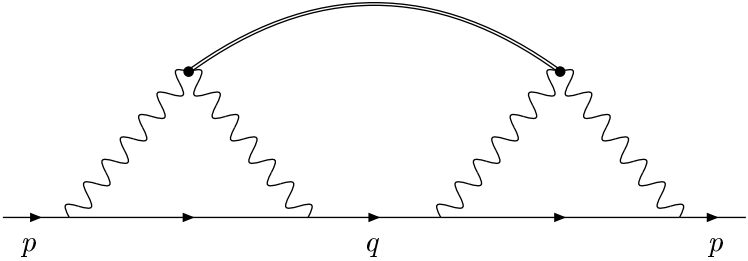}
\caption{ The self-energy term of the mass gap equation} 
\label{massgapfig} 
\end{figure}

The dependence in the relative momentum must comply with the linear confinement which
is derived from the string picture,
\begin{equation}
-i \, V( {\bf p}-{\bf q} )=-i \, { \sigma \over {\cal C}}{ - 8 \pi \over |{\bf p}-{\bf q} |^4}
\end{equation}
where $\sigma \simeq $ is the string constant, and ${\cal C}$ is an algebraic color 
factor.

The gluon propagators and the different sub-vertices result in a distribution in the loop
momentum $k$. Here different choices would be possible. For simplicity we assume that
the relative momentum $p-q$ flows equally in the two effective gluon lines. We also
remark that the distribution is $k$ is normalized to unity once the correct string tension 
is included in the relative potential $V(p-q)$. This amounts to consider that the momentum
$k$ distribution is a Dirac delta,
\begin{equation}
(2 \pi)^3 \delta^3 \left({\bf k} -{ {\bf p} + {\bf q} \over 2}\right) \ .
\end{equation}

We finally compute the vertex, decomposing the Dirac propagator in the convenient 
particle and anti-particle propagators, computing the energy loop integral, and 
summing in color indices,
\begin{equation}
{\cal V}_{eff} = \lambda^c \ 
(S_k + C_k\, \vers{k}\cdot\trD{\gamma}) \ 
\Biggr|_{k={p+q \over 2}} \ , 
\label{vertex result}
\end{equation}
where $S_k=m_k/\sqrt{k^2+{m_k}^2}, \
C_k=k/\sqrt{k^2+{m_k}^2}$, 
and where $m_k$ is the constituent quark mass, to be determined in the next section

\section{Mass gap equation}
\label{sec:mass gap}

\begin{figure} 
\includegraphics{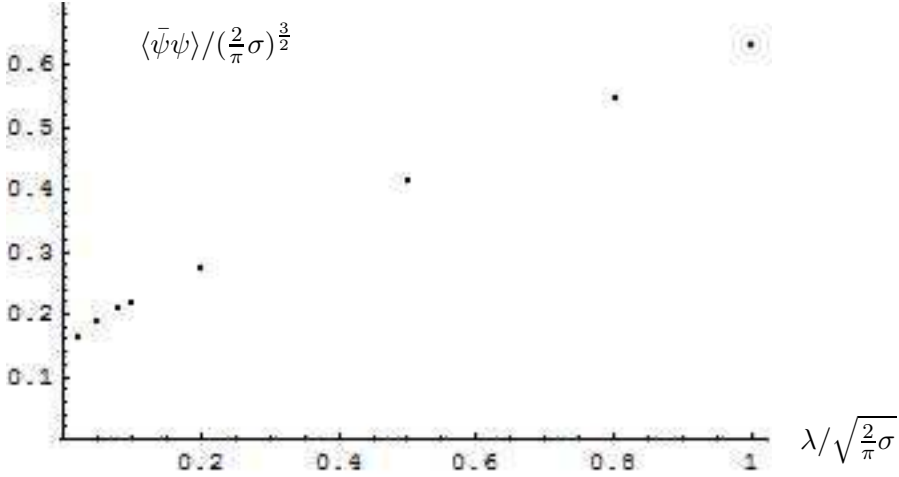}
\caption{ Testing the convergence of the numerical method with the quark
condensate $\langle \bar \psi \psi \rangle$.} 
\label{testing in condensate} 
\begin{picture}(0,0)(0,0)
\put(50,200){$\langle \bar \psi \psi \rangle /({2\over \pi}\sigma)^{3\over 2}$}
\put(300,50){$\lambda /\sqrt{{2\over \pi}\sigma}$}
\end{picture}
\end{figure}

We derive the mass gap equation projecting the Schwinger-Dyson equation with Dirac spinors,
\begin{equation}
\spub{s}{p}{\cal S}_0^{-1}(p)\spv{s'}{p} - \spub{s}{p}\Sigma(p)\spv{s'}{p} =0
\label{mass gap eq}
\end{equation}
where ${\cal S}_0$ is the free Dirac propagators, and where $\Sigma(p)$ 
is the self-energy, depicted in Fig. \ref{massgapfig}. The self energy consits
in a three loop Feynman diagram, including one loop in each double vertex and
a third rainbow-like loop for the string exchange interaction.
Nevertheless each double vertex is simple, see eq. (\ref{vertex result}), and we get
for the self-energy term, 
\begin{eqnarray}
\spub{s}{p}\Sigma(p)\spv{s'}{p} &=& \hspace{3mm}\int\frac{d^3q}{(2\pi)^3}\left[
(C_k^2-S_k^2)S_q C_p + 2S_k C_k S_q S_p\vers{k}\cdot\vers{p} - \right. 
\nonumber\\ 
&& - C_q S_p\vers{q}\cdot\vers{p} - 2S_k C_k C_q C_p \vers{k}\cdot\vers{q} + 
 \left. + 2C_k^2 C_q S_p \vers{k}\cdot\vers{q}\ \vers{k}\cdot\vers{p}
\right]V\left(|\trD{p}-\trD{q}|\right) \ .
\label{self energy eq}
\end{eqnarray}

The mass gap equation is a difficult non-linear integral equation,
that does not converge with the usual methods
\cite{Yaouanc,Adler,Bicudo.solutions}. Here we develop a method to solve
it with a differential equation, using a convergence parameter 
$\lambda \rightarrow 0$. 
Our technique consists in starting with a large infrared cutoff $\lambda$,
where the integral term in eq. (\ref{self energy eq}) is small. Then
eq. (\ref{mass gap eq}) for the chiral angle $\varphi_p$ is essentially 
a differential equation which can be solved with the standard shooting method
\cite{Bicudo.solutions}. 
Next we decrease step by step the $\lambda$ parameter, using as an
initial guess for the evaluation of the integral the $\varphi_p$ determined
for the previous value of $\lambda$. 
We test the convergence of the 
method computing the quark condensate, $\langle \bar \psi \psi \rangle $,
see Fig. \ref{testing in condensate}.

\section{Results and conclusion}
\label{sec:conclusion}

\begin{figure} 
\includegraphics{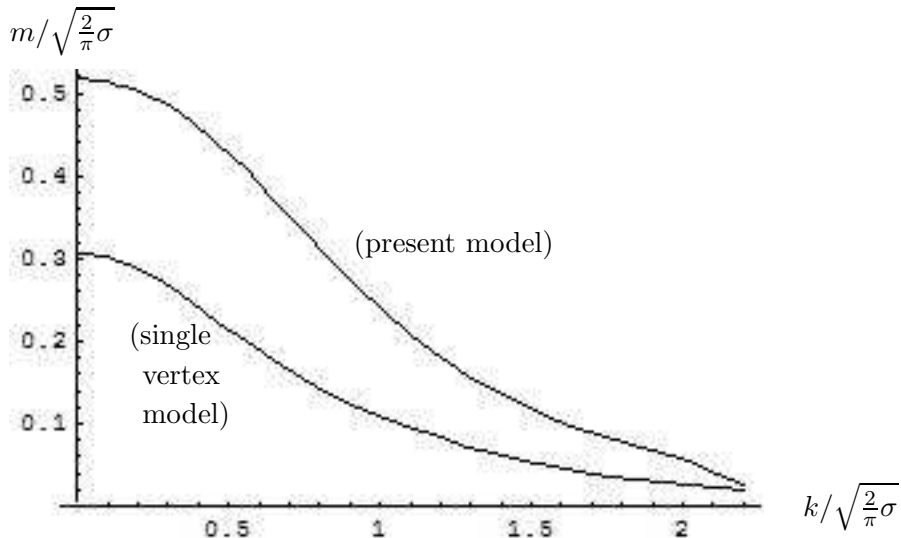}
\caption{ The $m_k$ solutions of the mass gap equation in units of 
$\sqrt{{2\over \pi}\sigma}$.} 
\label{massgapsolved}
\begin{picture}(0,0)(0,0)
\put(130,150){(present model)}
\put(45,115){(single}
\put(50,100){vertex}
\put(50,85){model)}
\put(300,50){$k/\sqrt{{2\over \pi}\sigma}$}
\put(00,230){$m/\sqrt{{2\over \pi}\sigma}$}
\end{picture}
\end{figure}

In this talk we build a model for the coupling of quark to a scalar string. Double vector
vertices are used, and the quark confining interaction has a single parameter, the
string tension $\sigma\simeq 200MeV/Fm$.
We solve the mass gap equation for the spontaneous breaking of chiral symmetry.
In Fig. \ref{massgapsolved} we compare the constituent quark mass $m_k$, computed
with our double vertex defined in eq. (vertex result), with the mass computed
with a simple Coulomb gauge vertex $\lambda^c \, \gamma^0$. It turns out that
the dynamical quark mass $m_k$ is larger when the double vertex is used, and this
is a good point for the present work
\cite{Adler}.

In the chiral limit of a vanishing quark mass, the effective vertex 
(\ref{vertex result}) 
${\cal V}_{eff}\rightarrow \lambda^c \vers{k}\cdot\trD{\gamma}$ is
proportional to the $\gamma^\mu$ and is 
therefore chiral invariant as it should be, whereas in the heavy quark limit,
 ${\cal V}_{eff}\rightarrow \lambda^c $ is simply a scalar vertex.
The dynamical generation of a quark 
mass $m_k$ also generates a scalar coupling for light quarks.

The results are encouraging, and we will now try to reproduce the whole hadron 
spectrum.

\end{document}